\documentclass[prb,twocolumn,showkeys,preprintnumbers,amsmath,amssymb]{revtex4}

\usepackage{graphicx}
\usepackage{dcolumn}
\usepackage{bm}

\begin{document}

\title{Cryogenic Sapphire Oscillator using a low-vibration design pulse-tube cryocooler: First results.}

\author{John G.~Hartnett,$^1$ Nitin R.~Nand,$^1$ Chao~Wang,$^2$ Jean-Michel~Le Floch$^1$}
 \email{john@physics.uwa.edu.au}
\affiliation{$^1$School of Physics, University of Western Australia 35 Stirling Hwy Crawley 6009 W.A. Australia\\$^2$Cryomech, Inc., 113 Falso Drive, Syracuse, NY 13211, USA}

\date{\today}

\begin{abstract}
A Cryogenic Sapphire Oscillator has been implemented at 11.2 GHz using a low-vibration design pulse-tube cryocooler.  Compared with a state-of-the-art liquid helium cooled CSO in the same laboratory, the square root Allan variance of their combined fractional frequency instability is $\sigma_y = 1.4 \times  10^{-15}\tau^{-1/2}$ for integration times $1 < \tau < 10$ s, dominated by white frequency noise. The minimum $\sigma_y = 5.3 \times 10^{-16}$ for the two oscillators was reached at $\tau = 20$ s. Assuming equal contributions from both CSOs, the single oscillator phase noise $S_{\phi} \approx -96 \; dB \; rad^2/Hz$ at 1 Hz offset from the carrier.
\end{abstract}


\maketitle

\section{Introduction}
Cryogenic sapphire oscillators (CSOs) have been under development at the University of Western Australia for about the last two decades \cite{Locke2008} and have been shown to have state-of-the-art performance \cite{Hartnett2006}.  Sapphire has extremely low loss at microwave frequencies \cite{Braginsky} and lends itself perfectly to the development of these microwave oscillators. The oscillators are operated with a resonator tuned to a Whispering Gallery mode with high order azimuthal mode number, which gives them large immunity to environmental effects. The CSOs have been deployed to a number of frequency standards labs \cite{Watabe2006, Watabe2007} and have facilitated atomic fountain clocks  reaching a performance limited only by quantum projection noise \cite{Santarelli1999}.  They have been used for fundamental tests of physics \cite{Wolf2003, Stanwix2005, Stanwix2006}.  Similar oscillators have been developed at JPL \cite{Dick1, Dick2, Wang} where application to deep space tracking was their ultimate goal. 

This current project has been to develop a local oscillator for radio astronomy, particularly VLBI, where an improved short-term stability of about two orders of magnitude over the hydrogen maser offers significant gains. Due to the high altitude of very high frequencies VLBI sites it may be that the quality of the signal is limited by the local oscillator. If it could be demonstrated that VLBI signal integration could be carried out over shorter time intervals hence mitigate against atmospheric decoherence, this may produce clearer images.  Particularly, the advantage of using this technology may be found in the development of the Square Kilometer Array (SKA) radio-telescope, which will rely on widely separated antenna sites.

\begin{figure}[!t]
\centering
\includegraphics[width=3.5in]{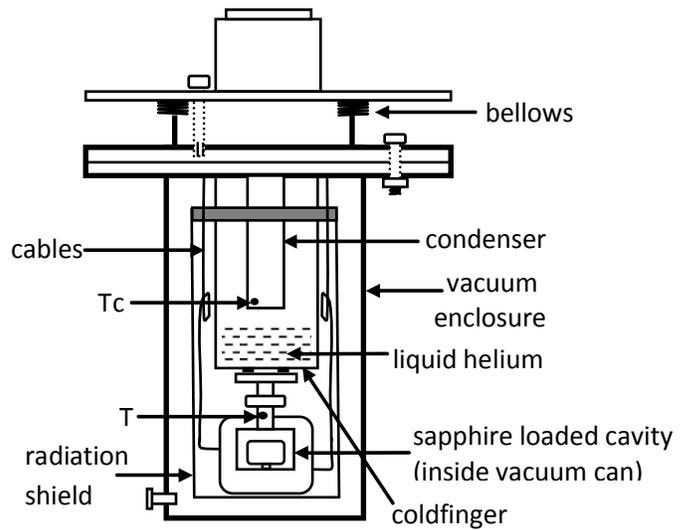}
\caption{Schematic of cryocooler cryostat showing vacuum enclosure, the bellows and the location of sapphire loaded cavity on the `coldfinger' that is cooled by a few liters of liquid helium maintained by the condenser inside the helium gas space. The location of the cavity ($T$) and condenser ($Tc$) control points are indicated.}
\label{fig_1}
\end{figure}

\section{New design with cryocooler}
A highly stable CSO, cooled with a 2-stage CryoMech PT407 low-vibration pulse-tube cryocooler, has been constructed with a cylindrical resonator made from Crystal Systems HEMEX grade \cite{CS} single-crystal sapphire. The new cryocooled CSO contains a nominally identical resonator to those in the liquid helium cooled CSOs used in the lab. The design of sapphire resonators used in these CSOs has been previously discussed \cite{Tobar2006} as has the loop oscillator and its servo control systems \cite{Hartnett2006, Locke2008}. 

The operational mode  was chosen to be the $WGH_{16,0,0}$ mode with a frequency of 11.202 GHz. The mode nomenclature means it has 16 azimuthal variations in the electromagnetic field standing wave around half its circumference, and one variation each along the radial and axial axes.  The new resonator exhibited a turning point in its frequency-temperature dependence at about 5.984 K and has a loaded Q-factor of $1.05 \times 10^9$ at the turnover temperature. The resonator primary and secondary port coupling coefficients at the turnover temperature were set to 0.80 and about 0.01, respectively.

The microwave energy is coupled into the resonator with loop antenna probes, through the lateral cavity wall, made from the same coaxial cables that connect it to the loop oscillator in the room temperature environment. Using an Endwave JCA812-5001 amplifier with sufficient microwave gain (nearly 50 dB), a microwave filter set at the resonance frequency and the correct loop phase, set via a mechanical phase shifter, we get sustained oscillation. 

The details of the loop oscillator, the temperature control of the sapphire loaded cavity, the power control of the loop oscillator and the Pound frequency servo locking the oscillator frequency to the resonance in the sapphire crystal are unchanged in the new design. The difference is that the new oscillator has a significantly different cryostat housing the vacuum can enclosing the sapphire loaded cavity. See Fig. 1 for a schematic. The thermal design of the cryostat and operation with the low-vibration cryocooler is discussed elsewhere \cite{Chao}. 

In the experiments reported here a liquid helium cooled CSO \cite{Hartnett2006}, operating on the same $WGH_{16,0,0}$ mode with a frequency of 11.200 GHz, was used to make frequency comparisons with the new cryocooled CSO. 

\begin{figure}[!t]
\centering
\includegraphics[width=3.5in]{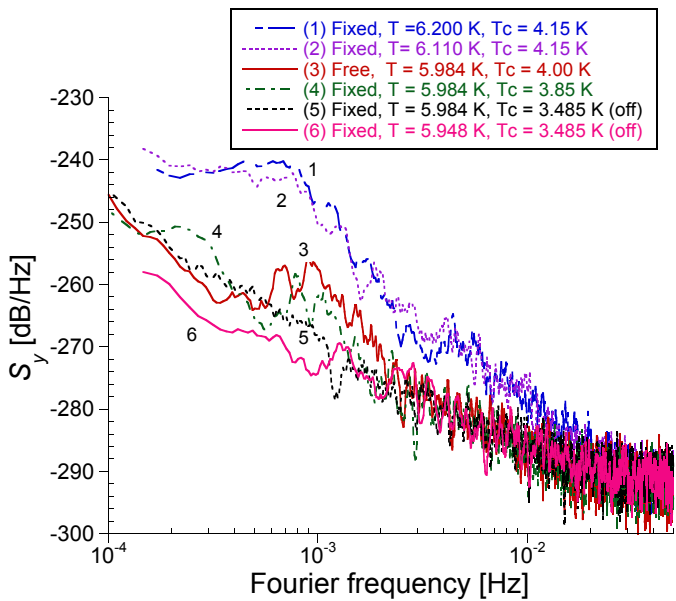}
\caption{(color online) $S_y \; [dB/Hz]$ calculated from the power spectral density of beat frequency data from time domain measurements using a $\Lambda$-counter with gate times of 10 s. Various experiments were performed operating at different control temperatures on the cavity ($T$) and on the condenser ($Tc$). Fixed and Free indicate whether the bellows was held fixed by three securing bolts or allowed to be free.}
\label{fig_2}
\end{figure}
\begin{figure}[!t]
\centering
\includegraphics[width=3.5in]{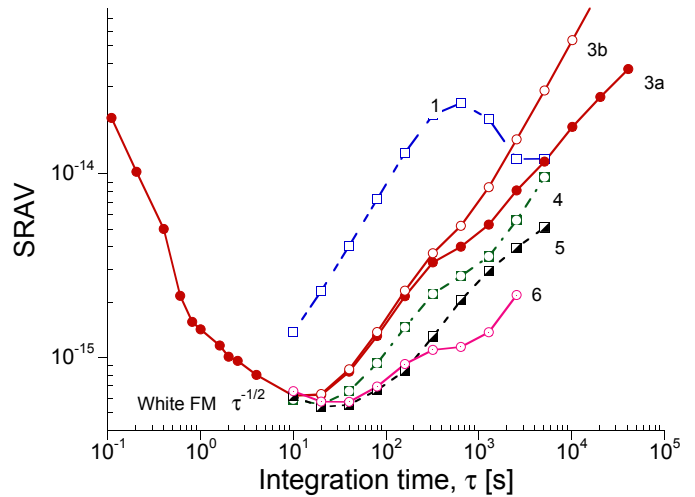}
\caption{(color online) Square root Allan variance (SRAV) $\sigma_y$ calculated from the same time domain beat data using a $\Lambda$-counter with gate times of 10 s as used in Figure 2. Additional measurements were made for $\tau \leq 1$ s and are shown on line (3). See text for details. All data runs with a gate time of 10 s were for durations of many hours, generally overnight. To display the results clearly error bars are not shown, but for $\tau \leq 100$ s the error bars are smaller than the markers used. }
\label{fig_3}
\end{figure}

The new cryocooler cryostat has three main design features worth noting. The cold-head motor has been removed from the cold head, which is not shown in Fig. 1. The cryostat contains a cylindrical space where the condenser liquefies a few liters of pure helium gas (labeled ``liquid helium'' in Fig. 1), and finally this gas region has flexible stainless steel bellows which permit it to be operated in either the condition with the bellows fixed with bolts or free to move. The latter was designed primarily to allow operation with positive pressure in the helium gas space. No rigid links exist between the ``condenser'' of the cryocooler and resonator. The principle is that by separating the condenser, which has about 12 $\mu m$ vertical motion at approximately 1.4 Hz, from the point where the experiment is attached (labeled ``coldfinger'' in Fig. 1) we attenuate vibration induced frequency shifts in the resonator. 

\section{Stability}
A number of experiments were run with different control temperatures  on the copper post ($T$ in Figs 2 and 3) supporting the sapphire loaded cavity and with different control temperatures on the cryocooler condenser ($Tc$ in Figs 2 and 3) inside the helium gas region. The resonator frequency-temperature turnover point was determined from measuring $T$ on the copper post. When $Tc < 4.2$ K a vacuum develops inside the helium gas region. Curves (5) and (6) resulted when  temperature control there was switched off. Note: The curves are numbered in Figs 2 and 3 where they correspond, i.e. the same data is used to generate the like numbered curves.

Fig. 2 shows $S_y \; [dB/Hz]$ calculated from the power spectral density of fractional frequency measurements of the 1.58 MHz beat between the two CSOs measured with an Agilent 53132A $\Lambda$-counter \cite{Dawkins} with a 10 s gate time. (When using this type of counter the calculated SRAV results differ slightly from those calculated from a standard counter. The reader is advised to refer to Ref. \cite{Dawkins} where a full analysis is given.) Six individual experiments are listed in the figure caption. Curves (1) and (2) are where the temperature was deliberately set above the frequency turnover point, to enhance any temperature induced frequency fluctuations. A significant hump in the noise is seen centered around a Fourier frequency $f = 10^{-3}$ Hz. The data of curve (1) is shown in Fig. 3 as SRAV ($\sigma_y$) with a very significant corresponding hump at $\tau = 10^3$ s. All numbered curves in Fig. 3 where $\tau \geq 10$ s corresponds to data, and the legend,  in Fig. 2. For all measurements where $\tau < 10$ s, the gate time of the measurements are the same as the indicated points. There were only very marginal differences in $\sigma_y$ for $\tau < 1$ s, for the experiments shown. 

\begin{figure}[!t]
\centering
\includegraphics[width=3.5in]{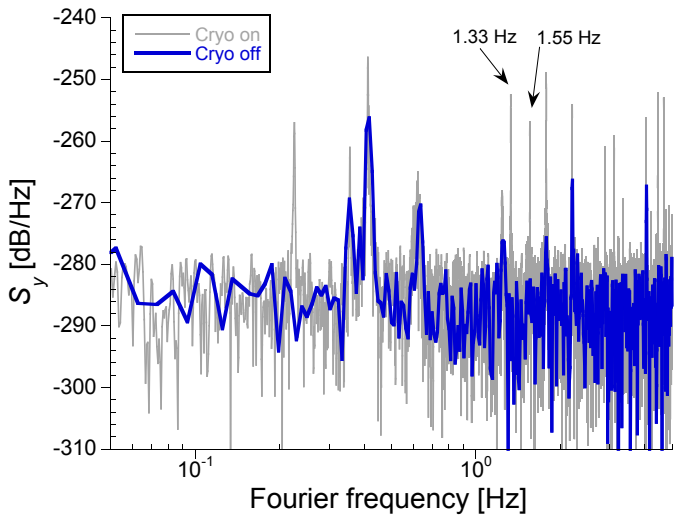}
\caption{(color online) $S_y \; [dB/Hz]$ calculated from power spectral density of time domain beat data using a $\Lambda$-counter, with a gate time of 0.1 s, where the data was recorded initially with the cryocooler running and then after it was switched off.  }
\label{fig_4}
\end{figure}
\begin{figure}[!t]
\centering
\includegraphics[width=3.5in]{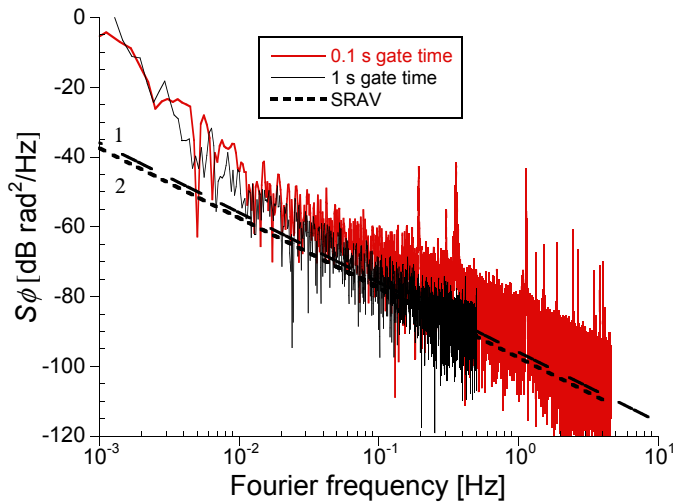}
\caption{(color online) $S_{\phi} \; [dB \; rad^2/Hz]$ calculated from power spectral density of time domain beat data using a $\Lambda$-counter, with gate times of 0.1 s and 1 s. 3 dB has been subtracted to represent the phase noise of a single CSO. The dashed line (curve 1) indicates the best fit $f^{-2}$ dependence to $S_{\phi}$ data (from the 1 s gate time measurements) where $f$ is the Fourier frequency offset from the 11.2 GHz carrier. The broken line (curve 2) is $S_{\phi}$ determined from $\sigma_y$ of Fig. 3 assuming white FM noise. }
\label{fig_5}
\end{figure}

At the turnover point $T = 5.984$ K, curves (3) and (4), in Fig. 2, show the effect of both fixed and free bellows, though the helium gas pressure was slightly lower in the case of curve (4) than in curve (3). Curve (4) shows a reduced hump in $S_y$ at $f = 10^{-3}$ Hz compared to curve (3). Compare their corresponding $\sigma_y$ curves (labeled (3) and (4)) in Fig. 3. A significant improvement can be seen. The bifurcation of $\sigma_y$ curves (3a) and (3b) in Fig. 3, shows the results when drift is removed. Curve (3b) contains the drift. Fractional frequency linear drift was measured at $3 \times 10^{-12}$/day and its source has not yet been determined. That will be the subject of future design improvements. In all other curves the drift has been removed.  Curves (5) and (6), in both Figs 2 and 3, result from data taken when the condenser temperature control was switched off and maximum vacuum (with a pressure of about 52 kPa) developed in the helium gas space. These conditions produced the best results. The hump at a Fourier frequency $f = 10^{-3}$ Hz vanished in Fig 2. In Fig. 3, it can be seen that operating slightly below what was initially determined to be turnover temperature (curve (6)) actually improved $\sigma_y$ in the longer term. This may be attributed to the fact that we approached closer to the correct turnover value for $T$.

\section{Phase noise}
The beat was also counted with a gate time of 0.1 s and converted to $S_y$, both when the cryocooler was running and switched off. This is shown in Fig. 4, where the cavity temperature was controlled at 5.984 K. Much less data were available to be taken after the cryocooler was switched off because temperature control could not be maintained very long.  The cryocooler pump frequency  was expected to be found around 1.4 Hz. The closest two peaks are indicated in the `Cryo on' data, which are not found in the `Cryo off' data. It is apparent however that the noise is white frequency (white FM) at Fourier frequencies shown. Therefore the rise in $\sigma_y$ for $\tau \leq 0.5$ s with a slope of order $\tau^{-3/2}$ is indicative of the bright lines and the significant dead time when $\tau < 1$ s and not enhancement in phase noise. 

The phase noise calculated from frequency counter measurements is shown in Fig. 5. Here we have converted the $S_y$ data determined from $\Lambda$-counter measurements to phase noise using,
\begin{equation} \label{Sy}
S_{\phi}(f)=\frac{\nu_0^2}{f^2} S_y(f),
\end{equation}
where $\nu_0$ is the microwave oscillator frequency. 

The power spectral densities of fractional frequency data for both 0.1 s and 1 s gate time beat data were used. Those were the same data used to determine the  $\sigma_y$ at $\tau = 0.1$ s and 1 s  in curve (3) of Fig. 3, respectively.  It is quite clear from Fig. 5 that the phase noise level at 0.1 s gate time is much higher than at 1 s gate time. This we believe is largely due to unaccounted for dead time. Nevertheless by curve fitting to the 1 s gate time phase noise data in Fig. 5 for Fourier frequencies  $0.1 \leq f \leq 0.5$ Hz  we get the dashed curve (labeled curve (1)). This yields $S_{\phi}(f) \approx -95.9 -20 log(f) \; [dB \; rad^2/Hz]$ for the phase noise on the 11.2 GHz carrier of a single CSO. Here we assume that both oscillators contribute equally. 

The phase noise may also be related \cite{Dawkins},
\begin{equation} \label{sigy}
S_{\phi}(f)=\frac{\nu_0^2}{f^2} \frac{3 \tau}{2}\sigma_y^2,
\end{equation}
where white FM noise is assumed. The best fit to the SRAV curve (3) of Fig. 3 where $0.8 \leq \tau \leq 10$ s with $\tau^{-1/2}$ dependence yields $\sigma_y (1 s)= 1.44 \times 10^{-15}$. Using this result and (\ref{sigy}) we calculate the expected $S_{\phi}(f) \approx -97.5 -20 log(f)\; [dB \; rad^2/Hz]$ for a single CSO and it is shown as the broken curve (labeled curve (2)) in Fig. 5. This is in good agreement with the result (curve (1)) from $S_y$ derived using the 1 s gate time data.

The extrapolation of $S_{\phi}(1Hz)$ from 0.1 s measurements is not valid as seen in Fig 5. However the extension of the $S_{\phi}(f)$ to $1 \;Hz$ is reasonable from the 1 s gate time data and agreement is found with a white FM noise expectation determined from the measured values for $\sigma_y$ with gate times $0.8 \leq \tau \leq 10$ s. To properly characterize the oscillators a zero beat phase noise measurement needs to be carried out. In the future we plan to due this. 

\section{Conclusion}
In summary, the SRAV of the two oscillators follows $\sigma_y = 1.4 \times  10^{-15}\tau^{-1/2}$ for integration times $1 < \tau < 10$ s, dominated by white frequency noise, where an additional factor of $\sqrt{2}$ needs to be taken out for a single oscillator assuming they equally contribute. A minimum $\sigma_y = 5.3 \times 10^{-16}$ was reached at $\tau = 20$ s where the stability reaches a flicker floor, then for $\tau > 40$ s it rises crossing $1.2 \times 10^{-15}$ at $\tau = 1000$ s, with a $\tau^{1/2}$ dependence in the long term. The double-sided phase noise has been estimated at approximately $-96 \; dB \; rad^2/Hz$ at 1 Hz from the carrier, which compares favorably with $-87.5 \; dB \; rad^2/Hz$ reported by Marra et al. \cite{Marra}.

The cryocooled CSO stability certainly meets the requirements of a flywheel oscillator for atomic fountain clocks and it means that the cryocooled CSO could be implemented as a local oscillator that is a few orders of magnitude more stable, at integration times between 1 and 10 s, than a hydrogen maser that is currently used for VLBI radioastronomy. At about 1000 s the stability of the cryocooled CSO becomes comparable to that of a hydrogen maser and currently suffers from a linear frequency drift about 3 orders of magnitude worse than that in a hydrogen maser.  This we believe results from thermal gradients and ongoing work is trying to address this now. The liquid helium cooled CSO has a comparable drift to a hydrogen maser \cite{Hartnett2006} hence there is room to improve in this version also. However, the cryocooled oscillator does not suffer from the need for reliable liquid helium supplies in remote sites, often far from liquefication plants.  The only requirement is that sufficient power is available to run the 3-phase compressor. And the cryocooler can operate from 3 to 5 years without maintenance. And even when a new condenser head is needed the changeover would only take a day at most.  The cryocooled CSO also has the potential to provide a much improved local oscillator for the future SKA telescope, which will have antenna sites located across vastly separated remote desert sites.

\section{Acknowledgments}
The authors would like to thank the Australian Research Council, Poseidon Scientific Instruments, the University of Western Australia, Curtin University of Technology and the CSIRO ATNF (Australian National Telescope Facility), who will provide a telescope site where the CSO will be tested as a local oscillator against a similar installation using a hydrogen maser. Also we thank E. Ivanov for useful discussions in regards to presenting the results.

\end{document}